\documentclass{article}

\usepackage[a4paper, margin=1in]{geometry}
\usepackage{graphicx} 
\usepackage{authblk}
\usepackage{algorithm}
\usepackage{algpseudocode}
\usepackage{makecell}
\usepackage{xcolor}
\usepackage{adjustbox}
\usepackage{array}
\usepackage{multirow}
\usepackage{tabularx}
\usepackage{comment}
\usepackage{amsmath}
\usepackage{appendix}
\usepackage{hyperref}
\DeclareUnicodeCharacter{2009}{,}

\begin{document}

\title{Quantum Computing Tools for Fast Detection of Gravitational Waves in the Context of LISA Space Mission}

\author[1,2]{Maria-Catalina Isfan}
\author[1]{Laurentiu-Ioan Caramete}
\author[1]{Ana Caramete \thanks{Corresponding author: Ana Caramete, acaramete@spacescience.}}
\author[1]{Daniel Tonoiu}
\author[1]{Alexandru Nicolin-\.{Z}aczek}

\affil[1]{Institute of Space Science – INFLPR Subsidiary, 409 Atomistilor, Magurele, Romania}
\affil[2]{Doctoral School of Physics, Faculty of Physics, University of Bucharest, 405 Atomistilor, Magurele, Romania}

\date{\today}

\maketitle

\begin{abstract}
    The field of gravitational wave (GW) detection is progressing rapidly, with several next-generation observatories on the horizon, including LISA. GW data is challenging to analyze due to highly variable signals shaped by source properties and the presence of complex noise. These factors emphasize the need for robust, advanced analysis tools. 
    In this context, we have initiated the development of a low-latency GW detection pipeline based on quantum neural networks (QNNs). Previously, we demonstrated that QNNs can recognize GWs simulated using post-Newtonian approximations in the Newtonian limit. We then extended this work using data from the LISA Consortium, training QNNs to distinguish between noisy GW signals and pure noise. Currently, we are evaluating performance on the Sangria LISA Data Challenge dataset and comparing it against classical methods.
    Our results show that QNNs can reliably distinguish GW signals embedded in noise, achieving classification accuracies above 98\%. Notably, our QNN identified 5 out of 6 mergers in the Sangria blind dataset. The remaining merger, characterized by the lowest amplitude, highlights an area for future improvement in model sensitivity. This can potentially be addressed using additional mock training datasets, which we are preparing, and by testing different QNN architectures and ansatzes.
\end{abstract}

\section{Introduction}

The detection of gravitational waves (GWs) is a rapidly evolving field, with several next-generation observatories—both space-based (e.g., LISA \cite{LISA}, AEDGE \cite{AEDGE}, DECIGO \cite{DECIGO}, Taiji \cite{Taiji}) and ground-based (e.g., Cosmic Explorer \cite{CosmicExplorer}, Einstein Telescope \cite{Einsteintelescope}, and upgrades to LIGO, Virgo, and KAGRA \cite{LIGOVIRGOKAGRA}). The LISA space mission \cite{LISA} is one of the largest and most important space missions that will be undertaken by ESA, in partnership with NASA in the coming years. LISA will detect GWs using time-delay interferometry. The low-latency analysis pipelines must process the data as quickly as possible, the requirement being that it has to do it within one hour from the time when the Level-0 (L0) data, that is raw spacecraft telemetry that has undergone basic processing such as de-multiplexing, time annotation, inclusion of quality flags and positioning information, and the correction or removal of corrupted items, is transmitted to the ground systems \cite{low_latency_colpi2024lisadefinitionstudyreport}.

GW signals, especially those within LISA's \cite{LISA} detection range, carry information about the origin and growth of the current massive black hole population, the stochastic GW backgrounds, the formation and evolution of compact binary stars, as well as the fundamental nature of gravity and the expansion of the Universe \cite{low_latency_colpi2024lisadefinitionstudyreport}. In this paper, we focus exclusively on GWs emitted by binary black holes (BBHs). GWs generated by BBHs have complex structures, their frequency and amplitude depending on a number of parameters of the source, such as masses, spins, eccentricity.
In addition, the majority of the white dwarf population in our Galaxy emits GWs that overlap and that are difficult to disentangle, acting as a source of noise \cite{binary_white_dwarfs_as_noise_10.1093/mnras/stz2834}, \cite{binary_white_dwarfs_as_noise_Peter_L_Bender_1997}, \cite{binary_white_dwarfs_as_noise_PhysRevD.104.043019}, \cite{binary_white_dwarfs_as_noise_PhysRevD.70.122002}. Although, tens of thousands of white dwarf binaries are expected to be resolvable \cite{binary_white_dwarfs_as_noise_10.1093/mnras/stz2834}, \cite{binary_white_dwarfs_as_noise_Peter_L_Bender_1997}.

There are many challenges involved in GW detection and data analysis \cite{GWdatanalysischallenges}. Given the diversity of sources, multiple detection pipelines are needed to cover them. Additionally, the possibility of overlapping signals complicates detection. The algorithms must be easily scalable to handle increasing volumes of data. GW detectors generate large, complex, and varied datasets. For instance, each LIGO detector records approximately 1 petabyte of raw data per year, which is reduced to about 20 terabytes of processed strain data per interferometer annually \cite{ligomanagementplan}. The complexity arises from the highly variable GW signals, whose phase and amplitude depend strongly on the properties of the source. Moreover, the presence of high level and intricate noise further complicates the analysis. Additionally, since each detector employs distinct detection methodologies, the resulting data structures differ significantly. These factors underscore the need for the development of advanced and robust data analysis tools. Given their long development timelines of such tools, it is also crucial to anticipate and integrate likely advances in computing software.

Recent literature shows advanced data analysis pipelines capable of simultaneously extracting and characterizing signals from both Galactic binaries and massive binary black holes (MBHBs), enhancing our ability to interpret LISA's future observations \cite{LISApipelines1} \cite{LISApipelines2} \cite{LISApipelines3} \cite{LISApipelines4} \cite{LISApipelines5}. Additionally, glitches present significant challenges in the analysis of LISA data. These artifacts, which can mimic or obscure genuine GW signals, necessitate advanced modeling and mitigation strategies to ensure accurate detection and characterization of astrophysical sources \cite{LISAglitches}.

Before LISA \cite{LISA} becomes operational, it is crucial to ensure that robust data analysis tools are available. The research community is already working on developing these tools, applying modern approaches such as machine learning (ML) \cite{LISApipelines2} \cite{LISApipelines4} \cite{LISApipelines5} and even quantum computing.

Quantum computing is a revolutionary approach to solving complex problems in fields such as finance, optimization, and data analysis \cite{qcandqi} \cite{qalgorithms}. Quantum algorithms can require fewer steps than classical ones to complete a given task. Moreover, quantum computers operate fundamentally differently from classical computers due to the quantum nature of their hardware \cite{qcinthenisqeraandbeyond}.
Quantum Machine Learning (QML) \cite{QML} is a broad class of computational models inspired by classical ML that promises to reveal several quantum advantages, such as exponential speed-up \cite{supervisedlearningwithqc}. Such improvements in data processing speed could be especially valuable for missions like LISA, which will require the rapid and efficient analysis of large volumes of complex data \cite{low_latency_colpi2024lisadefinitionstudyreport}. In recent years, interest in QML has increased, along with improvements in model performance, including time series analysis \cite{QMLstateoftheart}.
A subcategory of QML is quantum neural networks (QNNs), characterized by the application of quantum computing techniques to the principles of classical neural networks (NNs) \cite{quantumneuralcomputing}.
Recently, various QNN models have been developed for forecasting \cite{timeseriesforecasting} \cite{timeseriesforecasting2} \cite{timeseriesforecasting3}, prediction \cite{timeseriesprediction}, classification \cite{timeseriesclassifierwithamplitudeembedding}, anomaly detection \cite{timeseriesanomalydetection}.

In astrophysics, QNNs have been applied to various computational tasks, including pulsar classification \cite{pulsarclassificationusingQNNs}, stellar classification \cite{stellarclassification}, telescope image classification \cite{telescopeimageclassification}.
Regarding GWs data processing, several quantum algorithms were implemented for GWs matched filtering using Grover's Algorithm \cite{GWmatchedfilteringgrover} and Quantum Monte-Carlo integration and quantum amplitude amplification \cite{GWmatchfilteringmontecarlo}, Bayesian parameter inference \cite{GWparameterinference}, the encoding of GW signals into quantum states \cite{GWencodingintoquantumstate}. Additionally, a recent study demonstrated hybrid classical-quantum spectral decomposition for time series, including GW signals recorded by LIGO \cite{quantumspectraldecompositionGWLIGO}.

However, there are few studies in the literature that demonstrate the use of QNN models for GW data processing. There is one paper that presents glitch detection in signals detected by LIGO \cite{GWglitchdetection}. We demonstrated in our proof-of-concept paper \cite{proofofconcept} that it is feasible to classify GW signals simulated in the Newtonian quadrupolar approximation at the adiabatic limit using QNNs and classical-quantum hybrid neural networks. In this paper, we extend our investigation to the Sangria LISA Data Challenge, a more complex and realistic mock dataset. To our knowledge, there are currently no published quantum data processing approaches specifically targeting the LISA space mission.

As a use case of quantum computing tools for the analysis of astrophysical data from space missions, we present a QNN that is able to rapidly and accurately detect GW signatures in detected signals. We test our application using data generated by the LISA Data Challenge (LDC) Working Group \cite{LDC} and we compare its performance against its classical counterpart.

This paper is structured as follows. In Sec.\ref{sec:sangria}, we present the Sangria LISA Data Challenge. Next, in Sec.\ref{sec:methodology}, we describe how we pre-process the data and the workflow of our QNN. The results are presented in Sec.\ref{sec:results}, followed by the comparison with the classical approach of the Sangria LISA Data Challenge in Sec.\ref{sec:discussion}. Finally, we present our conclusions in Sec.\ref{sec:conclusions} and future work in Sec.\ref{sec:future}.

\section{The Sangria LISA Data Challenge} \label{sec:sangria}

LDC \cite{LDC} has launched several mock data analysis challenges to help the community prepare for efficiently analyzing future GW signals that will be recorded by the LISA space mission \cite{LISA}. The latest challenge is the LISA Data Challenge 2a, nicknamed Sangria \cite{sangria}. It was simulated using the LDC Working Group's waveform-generating software \cite{LDCsoftware}. The aim of this challenge is to address and resolve mild GW source confusion under idealized instrumental noise conditions.

\begin{figure}
    \centering
    \includegraphics[width=\linewidth]{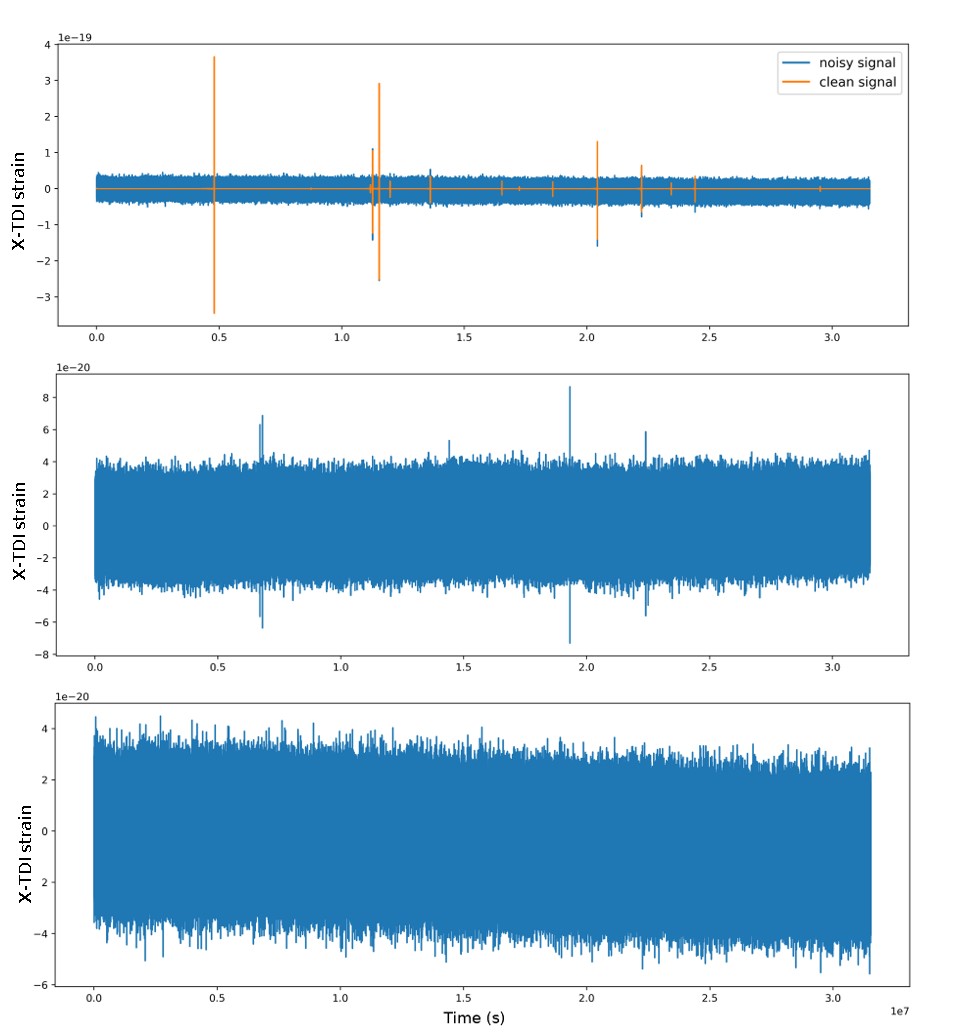}
    \caption{The training dataset for the Sangria Data Challenge, showing both the GW noisy signal and the GW noiseless signal (upper), the blind dataset for the Sangria Data Challenge (middle), and the noise of the Sangria training dataset, obtained by subtracting the clean signal from the noisy signal (lower). 'TDI' stands for Time Delay Interferometry, while 'X' represents the X projection of the signal}
    \label{fig:datasets}
\end{figure}

Sangria includes two datasets, each one under the form of an one-year time series:

\begin{enumerate}
    \item The training dataset, a dataset containing both noise and GWs signals (hereinafter referred to as 'noisy signal', where all the GW sources and their parameters are known; plus, the noiseless data (hereinafter referred to as 'clean signals') is provided (Fig.\ref{fig:datasets}, upper plot)
    \item The blinded dataset, which needs to be analyzed (Fig.\ref{fig:datasets}, middle plot), i.e. to find the number of GW sources and their kind. The number of sources and their kind are unknown.
\end{enumerate}

Both datasets contain merging MBHB sources exclusively, which were simulated using random source parameters.

The noise in both datasets is composed of Gaussian instrumental noise, verification galactic binaries noise, white dwarf binaries noise and stochastic background noise. The total noise can be isolated through subtracting the clean signal from the train dataset (Fig.\ref{fig:datasets}, lower plot).

Analyzing the Sangria LISA Data Challenge presents several difficulties. The main challenges are the following:

\begin{itemize}
    \item The noise amplitude is very high, many GW mergers are buried in noise.
    \item The GW sources are diverse, leading to significant variation in signal duration and merger amplitude.
    \item The training dataset is relatively short and limited, particularly for data analysis approaches such as machine learning. However, the adequacy of the dataset cannot be determined a priori; instead, it must be assessed heuristically through experimentation during model development and validation.
    \item In the feature space derived from signal processing, the training dataset and its corresponding noise are reasonably well separated, while the blind dataset's features tend to occupy the intermediate region between these two classes (Fig.\ref{fig:featssangria}).
\end{itemize}

\section{Methodology} \label{sec:methodology}

Our goal is to predict with high accuracy at what moments of time mergers occur in the Sangria blinded dataset. We will perform this data analysis using a QNN, based on the VQC algorithm. We will train it using the Sangria train dataset, a one-year time series containing 15 different GW signals immersed in noise.

\subsection{Data pre-processing} \label{sec:datapreprocessing}

The Sangria training dataset—with its 15 injected GW signals and their corresponding clean traces—and the separate blinded dataset each consist of a one-year time series sampled every five seconds. The data in this form is not enough to train and make predictions with a neural network (classical or quantum). Therefore, additional pre-processing steps must be applied before the data are fed into the neural network.

\begin{figure}
    \centering
    \includegraphics[width=\linewidth]{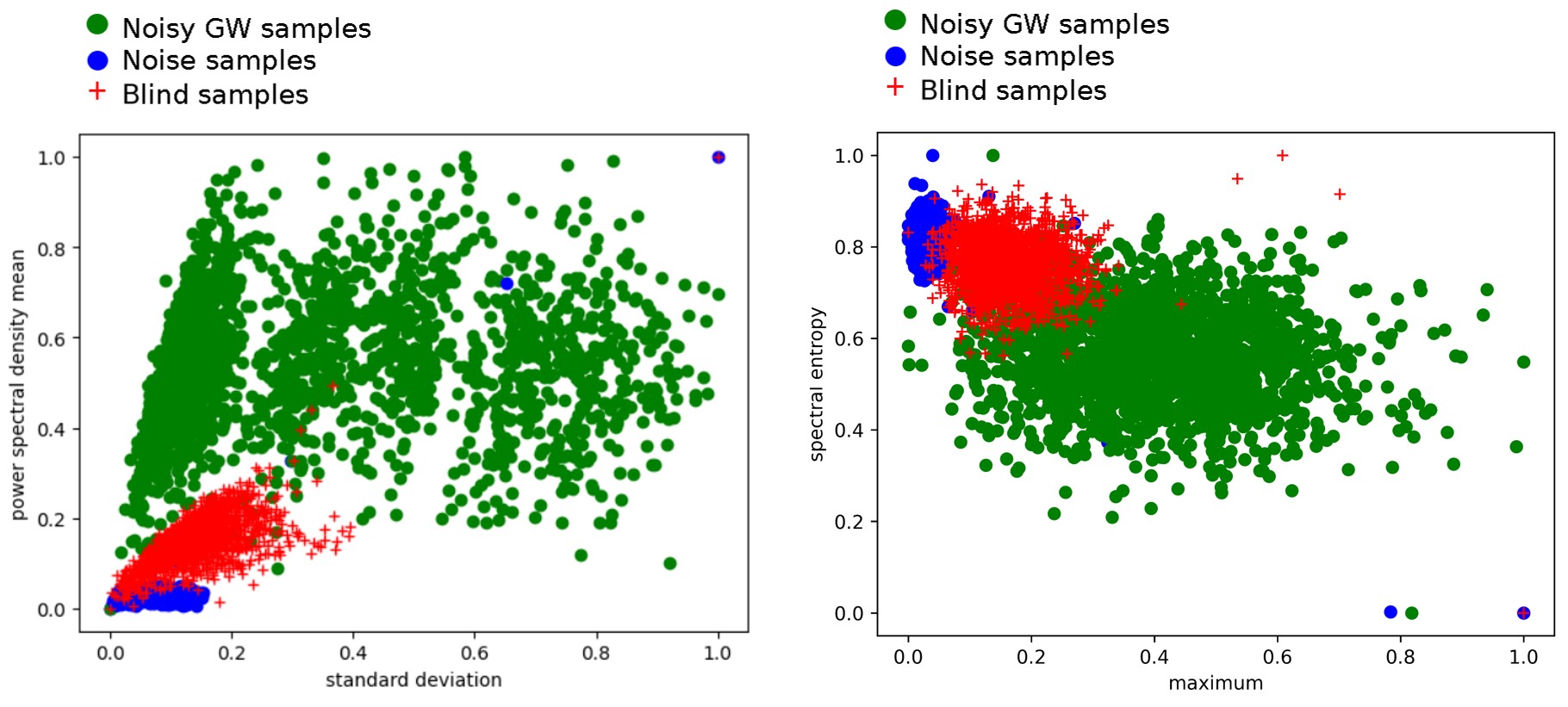}
    \caption{Power spectral density mean versus standard deviation (left) and spectral entropy versus maximum (right) for the noisy waveforms samples, respectively the noise samples in Sangria train and blind datasets. The samples were obtained using the procedure presented in Sec.\ref{sec:adjacentwindowsapproach}}.
    \label{fig:featssangria}
\end{figure}

\subsection{Data samples generation and labeling}

We divide the analysis of the Sangria blind dataset into two stages, following a warm-start training approach \cite{warmstartpretraining, warmstartinitialsolution, warmstartpretrainingfinetuning}:

\begin{enumerate}
    \item We first train the QNN to recognize the blind dataset as a GW, despite its high noise level.
    \item We then extract the optimized hyperparameters and the final state of the QNN’s trainable parameters, and use them to initialize the second stage of training.
\end{enumerate}

To obtain reliable results, a neural network requires a sufficiently large and diverse dataset. However, in this work, we are constrained to a single time series—the Sangria train dataset—containing only 15 distinct GWs. This limitation necessitates the implementation of additional sampling protocols to generate enough training examples for the QNN. For each training stage, we apply a different data sample generation and labeling strategy, described in Sections \ref{sec:adjacentwindowsapproach} and \ref{sec:movingwindowsapproach}. It is important to note that the two data splitting and labeling methods are mutually independent of each other.

\subsubsection{Adjacent windows sampling} \label{sec:adjacentwindowsapproach}

For the first training stage, we divide the Sangria train, noise and blind datasets into windows of 1802 adjacent samples. The number of windows was determined heuristically as defined in Eq.(\ref{eq:noofadjacentwindows}). Segments extracted from the Sangria dataset were assigned the label of '1' as they represent noisy GWs, whereas those from the noise dataset were labeled '0'. The QNN is trained on the Sangria and noise datasets. During inference, it is expected to assign the label '1' to all samples derived from the blind dataset, as this dataset contains only noisy GW samples.

\begin{equation}
    \text{no. of samples} = \left\lceil \frac{\text{no. of data points}}{\text{window length}} \right\rceil
    \label{eq:noofadjacentwindows}
\end{equation}

Using a window length of 3500, we generate approximately 1800 samples from each of the training and blind datasets, as well as from the noise signal extracted from the training dataset.

\subsubsection{Overlapping moving window sampling} \label{sec:movingwindowsapproach}

For the second training stage, we employ a moving window approach \cite{movingwindows1}, \cite{movingwindows2}. This method consists of passing a fixed size window over the data and selecting and storing the covered points at each moving step. Each subset of points obtained is treated as an individual sample. The dimension of the window has to be chosen, i.e., how many neighboring data points will be covered by a single window. The window starts at the first data point and moves forward one step along the data series until its end. The step size also must be chosen. Each subset of points covered by the window at every step is stored and treated as an individual sample.

We employ this moving window protocol to generate $\approx$63000 data samples from the Sangria train dataset and $\approx$63000 data samples from the blinded dataset, using a window length of 1000 data points and a step of 100 data points. The window length and the step size were set heuristically, according to Eq.(\ref{eq:noofmovingwindows}).

\begin{equation}
    \text{no. of samples} = \frac{\text{no. of data points} - \text{window length}}{\text{step}}
    \label{eq:noofmovingwindows}
\end{equation}

\begin{figure}
    \centering
    \includegraphics[width=\linewidth]{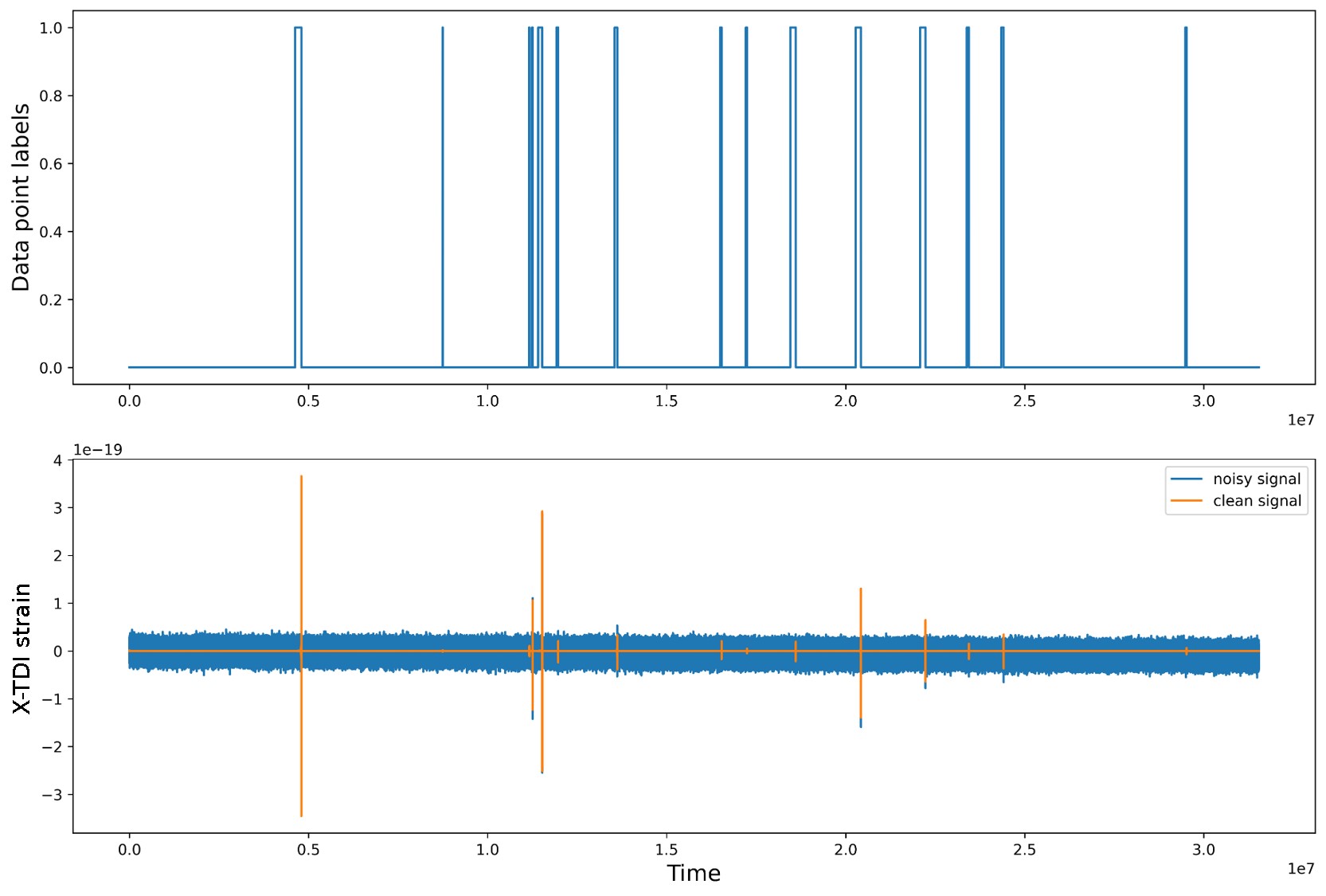}
    \caption{Assigned labels for each individual data point in the Sangria training dataset. This point-wise label list serves as the basis for assigning labels to samples generated via the moving window procedure.}
    \label{fig:datapointlabels}
\end{figure}

\begin{figure}
    \centering
    \includegraphics[width=0.7\linewidth]{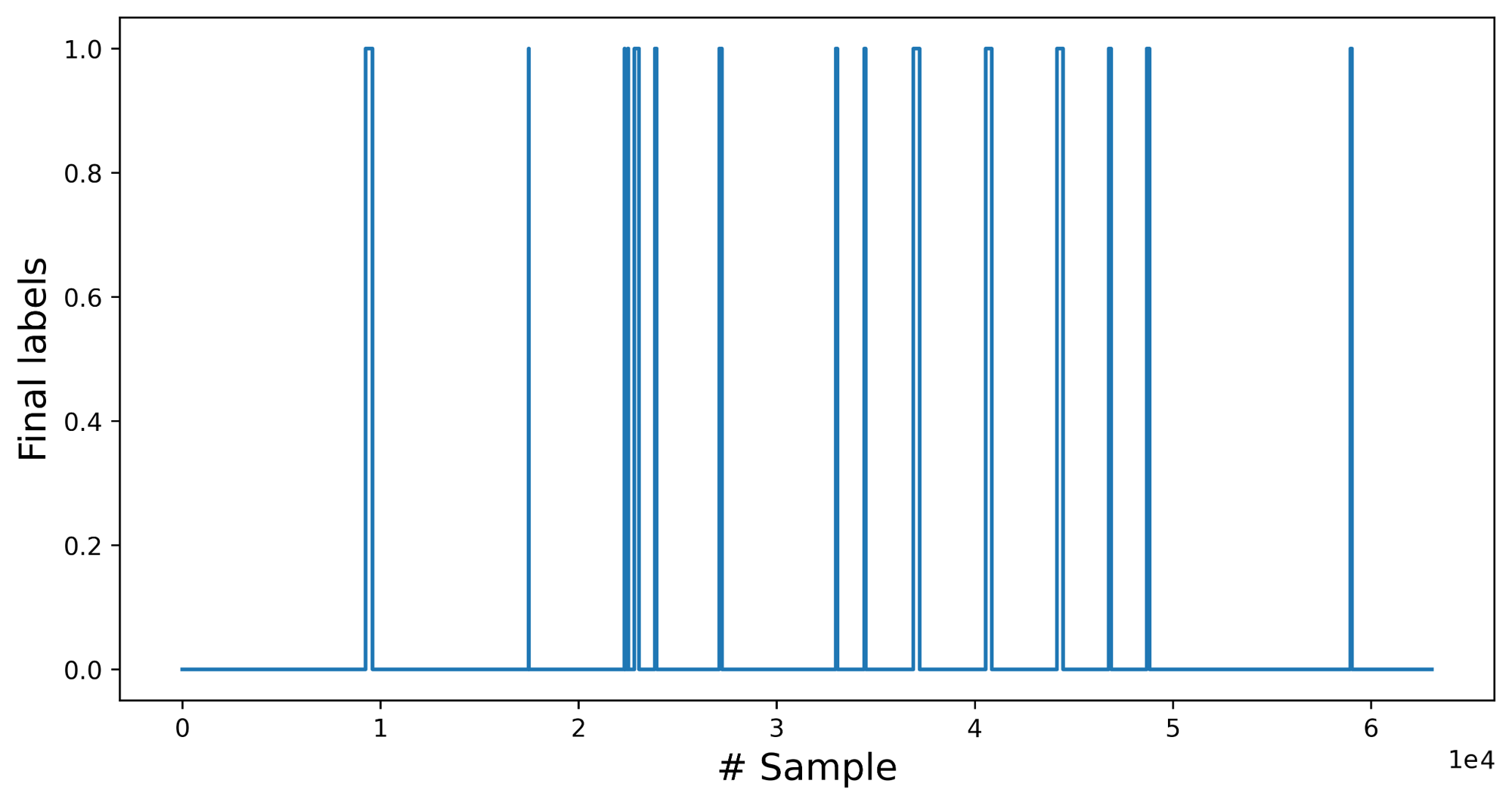}
    \caption{Final sample labels for the Sangria training dataset, used for the second training stage.}
    \label{fig:samplelabels}
\end{figure}

Each point of the training dataset consisting of 15 GW signals was labeled as follows: 1, if it fell within a window of approximately 4 days preceding and 27 minutes following the peak GW amplitude and 0 otherwise. This choice was motivated by the fact that in the absence of direct observational data on supermassive black hole mergers, the literature does not yet provide a precise timescale for the transition from inspiral to merger. Also, recent studies of black-hole binary mergers, such as Baker et al. (2008) \cite{Baker_2008} and Centrella et al. (2010) \cite{Centrella_2010}, support the above arguments by quantifying the waveform characteristics and emphasising the dominance of the late inspiral through ringdown signal. These works (and references therein) confirm that the final merger phase is brief in a relative sense but carries critical information and power, especially for the supermassive binary black holes targeted by LISA \cite{Centrella_2010}. Our selection of a merger window of 4 days pre-peak and 27-minute post-peak and inspiral otherwise is therefore well-founded in the current research landscape and should be enough to train the neural network to distinguish between merger and inspiral  (Fig.\ref{fig:datapointlabels}). This label list was then segmented into $\approx$63000 windows, consistent with the moving window procedure described above. Each window was assigned a label of '1' if it contained at least one data point labeled as '1'; otherwise, it was assigned a label of '0'. This process yielded the final window-level label list, also of length $\approx$63000 (Fig.\ref{fig:samplelabels}).

\subsubsection{Computation of normalized signal features for QNN input}

Our QNN takes classical input in the form of vectors, each vector element encoded into a separate qubit. Since simulating qubits on a classical computer is computationally expensive \cite{simulatingqubits1}, \cite{simulatingqubits2}, we compute compact feature vectors for each time series sample to keep the quantum input dimension manageable. By feature computation, GW signals are represented in a concise form that preserves essential characteristics for effective classification. Each qubit encodes a single feature.

Popular signal features are power spectral density mean, signal minimum, signal maximum, signal standard deviation, peak frequency, spectral entropy, 25th quantile, 50th quantile, 75th quantile \cite{signealprocessingfeatures}. We compute this set of features for each GW sample and perform feature-wise min-max normalization across the dataset, rescaling each feature independently to the $[0, 1]$ range. The resulting values are stored as normalized feature vectors. During QNN training, the optimal set of input features is selected heuristically, based on performance. After experimenting with different numbers of features, we found that using 4 features offered the best trade-off between model performance and simulation feasibility. As a result, our quantum circuit uses 4 qubits,

\subsection{The quantum circuit}

\begin{figure}[htbp]
    \centering
    \includegraphics[width=0.15\linewidth]{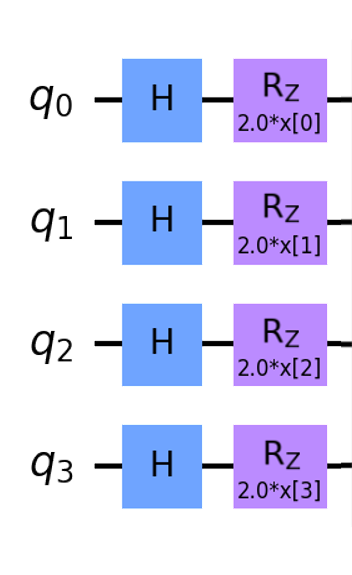}
    \caption{The feature map of the QNN: 1st order Pauli expansion circuit \cite{paulifeaturemap}. $x$ is the sample feature vector - classical input - that needs to be encoded in a quantum state with which the quantum program operates. Each qubit contains the information of one vector element $x[i], i=\overline{0,3}$.}
    \label{fig:ZFeatureMap}
\end{figure}

\begin{figure}[htbp]
    \centering
    \includegraphics[width=0.7\linewidth]{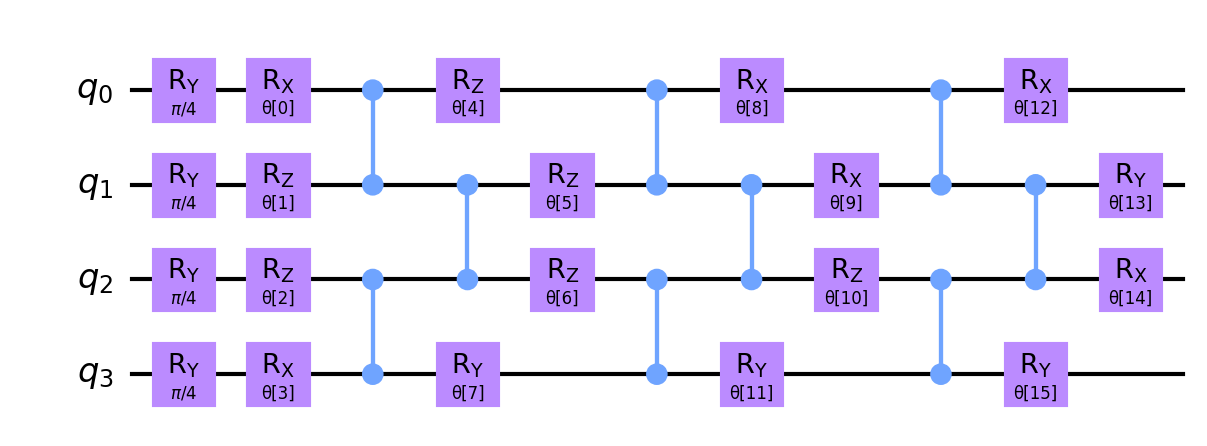}
    \caption{The ansatz of the QNN: Pauli two design circuit \cite{paulitwodesign1} \cite{paulitwodesign2}. The circuit is parameterized by the set of parameters $\theta[i], i=\overline{0,15}$. Within the QNN, the ansatz is repeated multiple times, with different parameters.}
    \label{fig:PauliTwoDesign}
\end{figure}

\begin{figure}[htbp]
    \centering
    \includegraphics[width=0.3\linewidth]{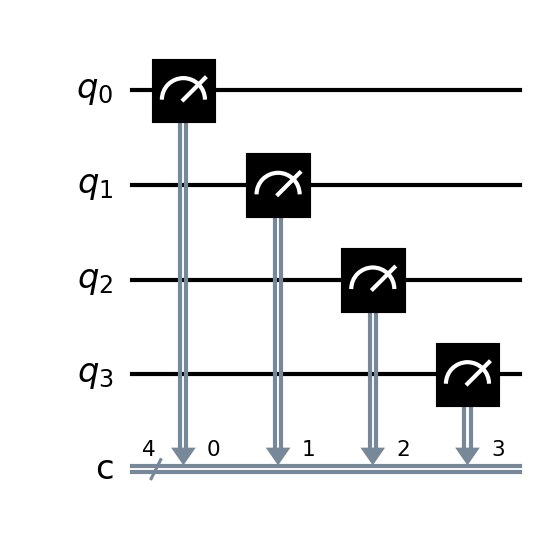}
    \caption{Measurement at the end of the quantum circuit. All qubits are measured.}
    \label{fig:measurement}
\end{figure}

The underlying quantum circuit of the QNN is made of three main blocks: the feature map \cite{featuremapref1} \cite{featuremapref2} (Fig.\ref{fig:ZFeatureMap}), which encodes each classical input sample into quantum information, the ansatz \cite{parameterizedquantumcircuitsasqmlmodels}(Fig.\ref{fig:PauliTwoDesign}), containing the trainable parameters, and the measurements (Fig.\ref{fig:measurement}). For further understanding of quantum computing basics, refer to Appendix \ref{app:basicqc}.

The feature map is a quantum circuit that has the role of encoding classical data into quantum data, with which the quantum computing algorithms process and operate with \cite{featuremapref1} \cite{featuremapref2}. This quantum data is represented by the output quantum state of the feature map. The feature map determines how the quantum program understands and interprets the classical data. The circuit involves a sequence of quantum gates that encode classical data points into the amplitudes and/or phases of the final quantum states. In our QNN circuit, the first-order Pauli-Z expansion circuit \cite{paulifeaturemap} plays the role of the feature map (Fig.\ref{fig:ZFeatureMap}).

The ansatz is a trainable quantum circuit that contains several trainable parameters (e.g. rotation angles of gates) per layer and has the role of mapping the input quantum encoded data into outputs \cite{parametizedquantumcircuits}. 
A QNN has several ansatz layers, each layer containing different parameter values \cite{parameterizedquantumcircuitsasqmlmodels}. The number of layers is chosen heuristically based on empirical performance of the QNN. All the layers determine the model and the architecture of the QNN. Our QNN's ansatz is the Pauli Two Design circuit \cite{paulitwodesign1} \cite{paulitwodesign2} (Fig.\ref{fig:PauliTwoDesign}). Each layer contains 16 trainable parameters.

In order to retrieve the output of a quantum computing program, some measurement is necessary at the end of the quantum circuit, followed by a specific post-processing protocol. In our QNN case, all qubit are measured in the end (Fig.\ref{fig:measurement}) in the computational basis $\{|0\rangle,|1\rangle\}$.

\begin{algorithm}
    \begin{algorithmic}[1]
        \State \textbf{Input:} Quantum circuit for some window
        \State Initialize $a \gets 0$ and $b \gets 0$
        \State Measure the circuit and obtain the bit strings probability distribution
        \For{each bit string}
            \State Compute the sum of the digits modulo 2
            \If{sum of the digits modulo 2 is 0}
                \State Add the sum to $a$
            \Else
                \State Add the sum to $b$
            \EndIf
        \EndFor
        \State The pair $[a,b]$ is now computed, where $a < 1$ and $b < 1$. $a$ represents the probability to predict the window label as '1' (merger) and $b$ represents the probability to predict the window label as '0' (non-merger)
        \If{a \textless b}
            \State Predict the label $[0,1]$ i.e. the sample is predicted as non-merger.
        \EndIf
        \If{a \textgreater b}
            \State Predict the label $[1,0]$ i.e. the sample is predicted as merger.
        \EndIf
    \end{algorithmic}
    \caption{} 
    \label{alg:measurementpostprocessing}
\end{algorithm}

\begin{figure}[htbp]
    \centering
    \includegraphics[width=\linewidth]{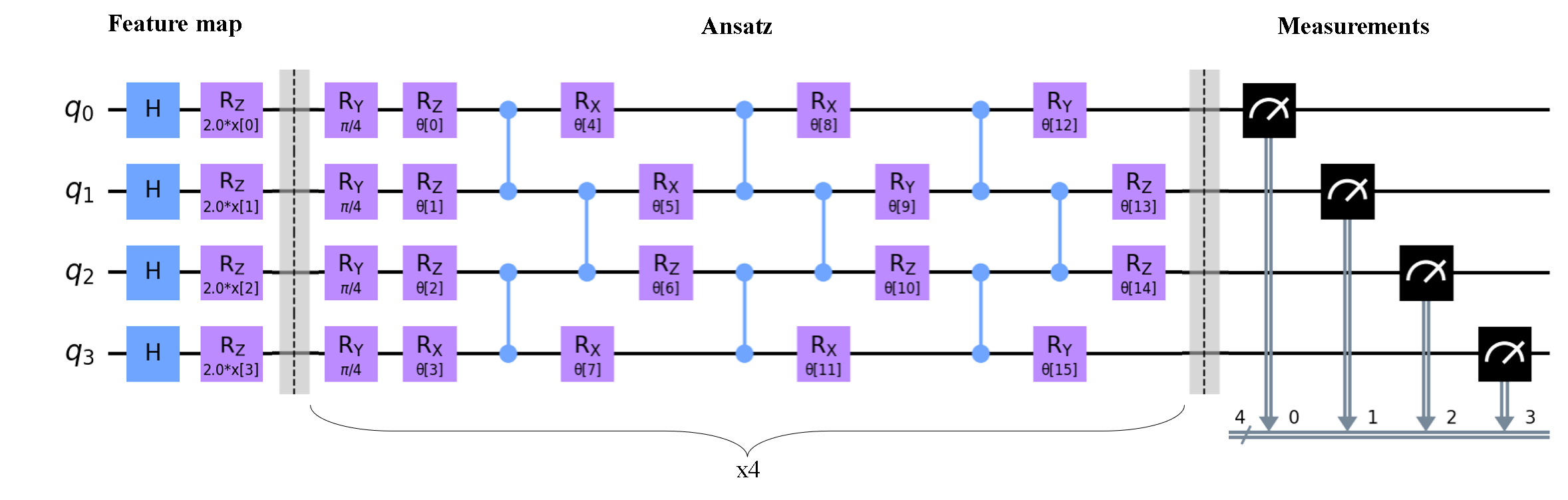}
    \caption{The full quantum circuit of the QNN. The ansatz is repeated four times.}
    \label{fig:QNNcircuit}
\end{figure}

The full quantum circuit of our QNN can be seen in Fig.\ref{fig:QNNcircuit}. The workflow is the following. Classical input enters the circuit as parameters of the feature map and is encoded in quantum input as the quantum state at end of the feature map. Then, this quantum state is processed by the layers of ansatz and the parameters are trained such that the final quantum state encodes the label of the input sample. Note that the ansatz is repeated four times, each time with different parameters, totaling 64 trainable parameters. Finally, the measurement is performed in the computational basis and following the measurement post-processing protocol in Algorithm \ref{alg:measurementpostprocessing} \cite{qiskit}, the predicted labels are obtained. The classical optimizer (COBYLA \cite{COBYLA}, in our case) compares the predicted labels with the true labels and tunes the ansatz parameters such that the loss function is minimized (binary cross entropy \cite{crossentropy}, in our case). For more details, see Appendix \ref{app:basicqnn}.

\subsection{The QNN structure}

\begin{figure}[htbp]
    \centering
    \includegraphics[width=0.8\linewidth]{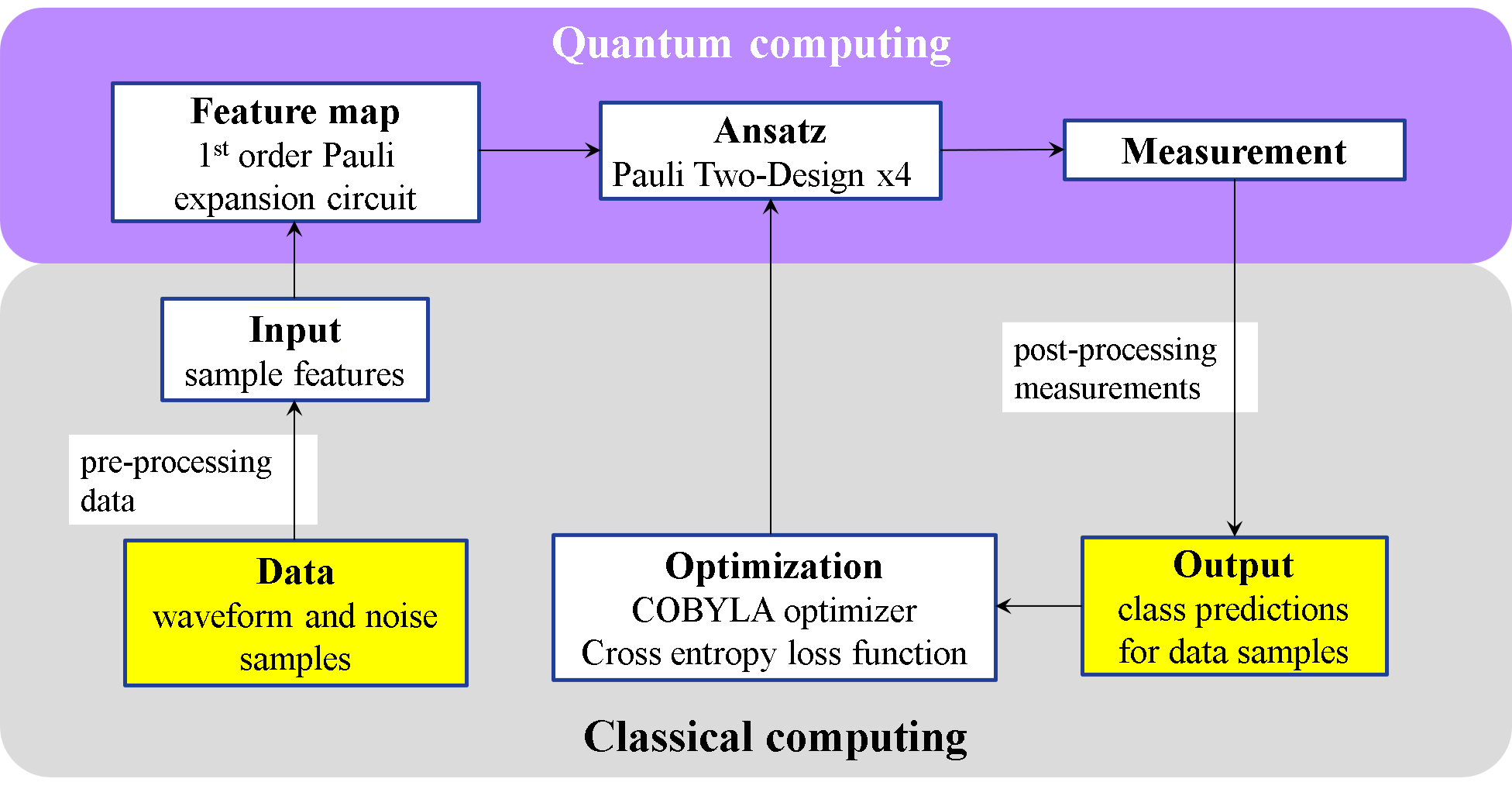}
    \caption{Algorithm structure of the QNN. The quantum part is highlighted in violet, and the classical part in gray. This algorithmic structure is the same for both training stages.}
    \label{fig:QNNalgorithm}
\end{figure}

The full algorithmic structure of the QNN is represented schematically in Fig.\ref{fig:QNNalgorithm}. It can be divided in two main blocks: the classical computing block and the quantum computing block. As it can be seen, our QNN is not a fully quantum program, as the pre-processing of the data, the optimization of the ansatz parameters and the measurement post-processing are implemented classically \cite{qiskit}. The model within the neural network and its architecture is quantum. See Appendix \ref{app:basicqnn} for more details.

The initial data consists of noisy GWs simulated using the LISA Data Analysis Software \cite{LDCsoftware}. Following the pre-processing described in Sec.\ref{sec:datapreprocessing}, the data is segmented into multiple samples, each converted into a feature vector input for the QNN. Each classical feature vector is encoded into quantum data via the feature map. The encoded quantum data is processed through layers of the ansatz, whose parameters are trained by the QNN. The quantum circuit’s final state is measured in the computational basis. Algorithm \ref{alg:measurementpostprocessing}is then applied to the measurement outputs to predict the label for each quantum data sample, corresponding to the input feature vectors.

\section{Results} \label{sec:results}

\begin{figure}[!htbp]
    \centering
    \includegraphics[width=0.6\linewidth]{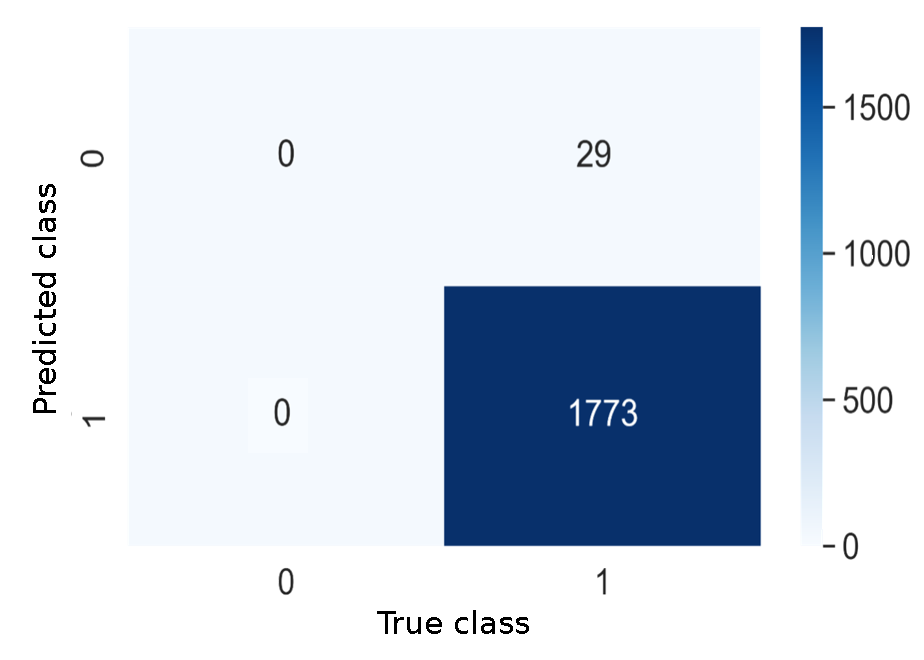}
    \caption{Confusion matrix for the QNN inference on the Sangria blind dataset.Most samples were correctly labeled as noisy GWs ('1'), with a small fraction misclassified as noise ('0').}
    \label{fig:result1}
\end{figure}

\begin{figure}[!htbp]
    \centering
    \includegraphics[width=\linewidth]{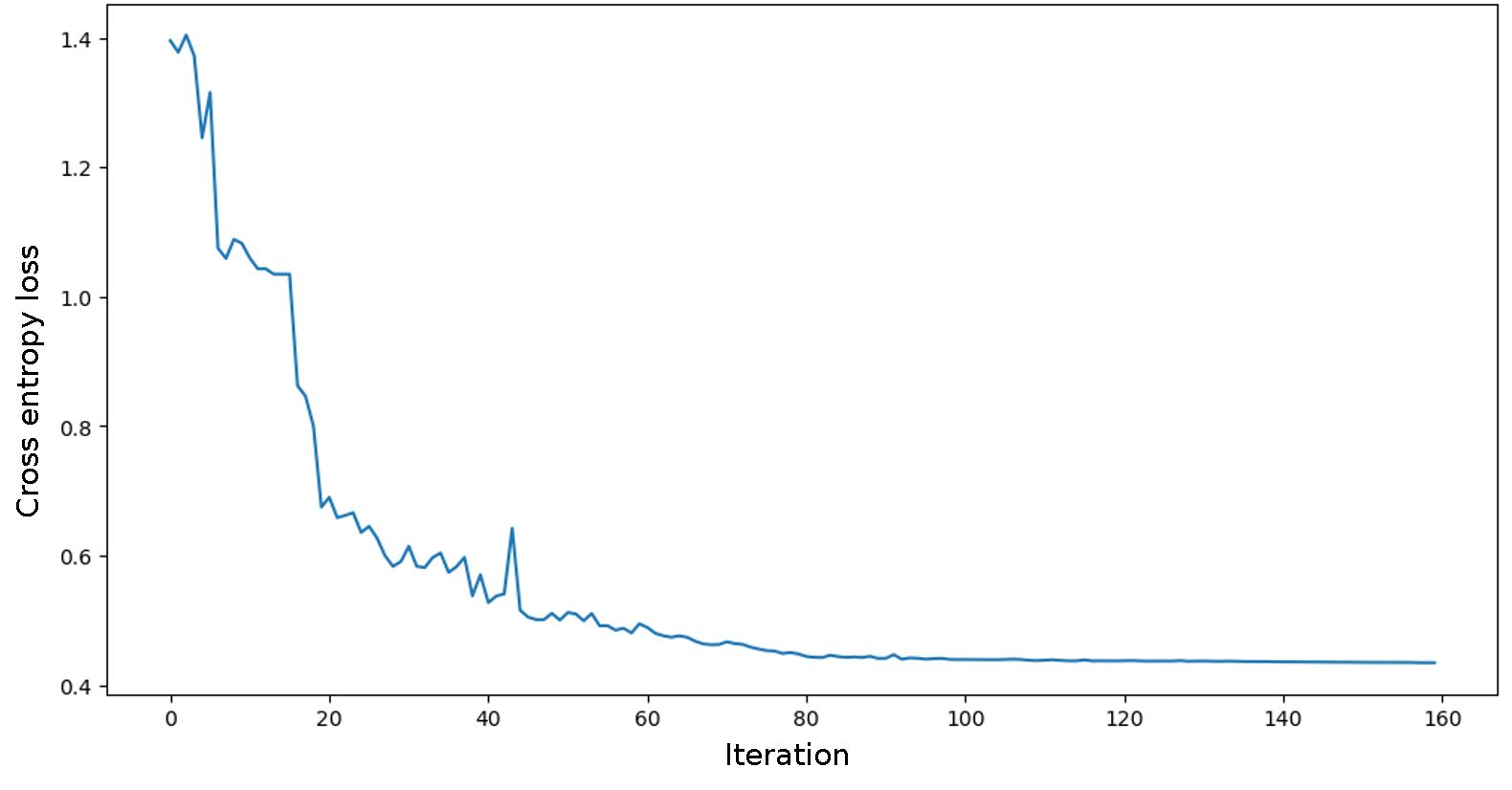}
    \caption{Training loss curve during classification of the Sangria blind dataset.}
    \label{fig:loss1}
\end{figure}

\begin{figure}[!htbp]
    \centering
    \includegraphics[width=0.6\linewidth]{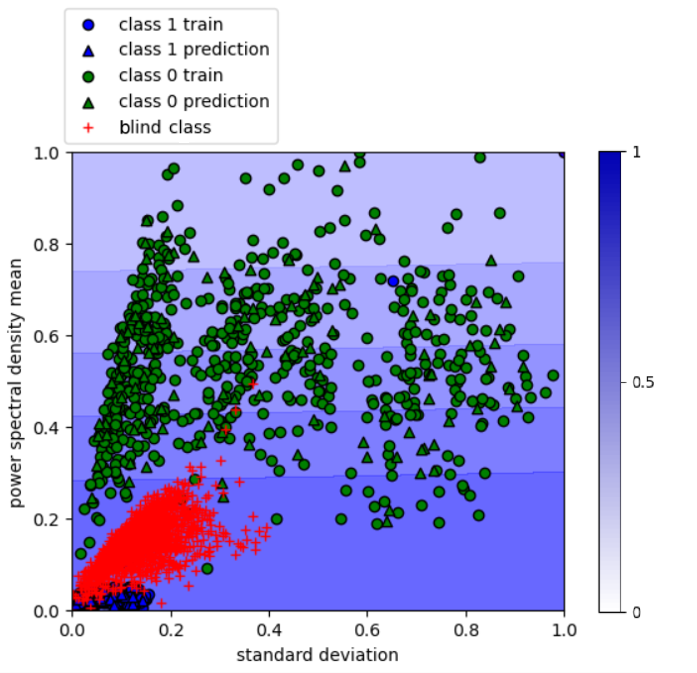}
    \caption{
    QNN decision regions after the first training stage, shown as a function of standard deviation versus power spectral density mean. The background gradient depicts the predicted probability of a noisy GW signal across the feature space.}
    \label{fig:decisionregionsgrad}
\end{figure}

Given the complexity and challenging nature of the datasets (Sec.\ref{sec:sangria}), before tackling the main objective of the Sangria LISA Data Challenge, we pre-trained our QNN to recognize the blind dataset as containing GW signals. For this task, we generated approximately 1800 adjacent data windows from each of the train dataset, the associated noise, and the blind dataset (Sec.\ref{sec:adjacentwindowsapproach}). Samples from the train dataset were labeled '1' (noisy GW), while noise samples were labeled '0' (noise). We then computed signal processing features as described in Sec.\ref{sec:datapreprocessing}. We used the warm start method \cite{warmstartincrementallearning} \cite{warmstartinitialsolution} \cite{warmstartpretraining} \cite{warmstartpretrainingfinetuning} during training to ‘teach’ the QNN to recognize the blind dataset as containing GW signals, i.e. to assign label '1' to all samples from the Sangria blind dataset. This strategy was essential for improving model performance, since the blind dataset lies between the GW and noise classes in the feature space (Fig.\ref{fig:featssangria}). Decision regions computed by the QNN after the first training stage (Fig.\ref{fig:decisionregionsgrad}) show a smooth gradient illustrating the variation in GW classification probability across the feature space. After training, predictions on the blind dataset yielded probabilities exceeding 98\% that these samples correspond to noisy GW signals (Fig.\ref{fig:result1}). The corresponding loss curve is shown in Fig.\ref{fig:loss1}.

\begin{figure}[!htbp]
    \centering
    \includegraphics[width=\linewidth]{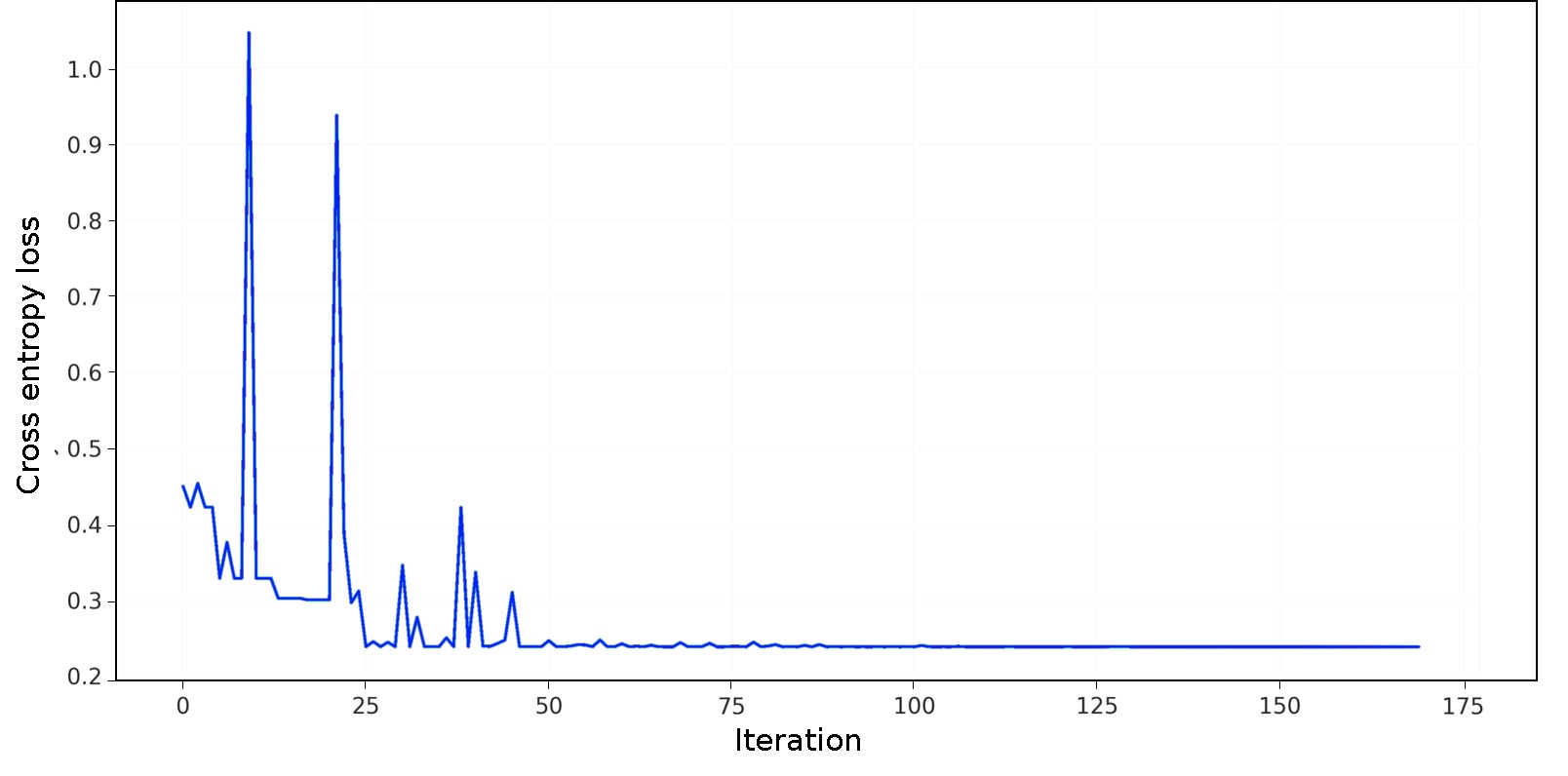}
    \caption{Training loss curve for GW merger detection in the Sangria blind dataset.}
    \label{fig:loss2}
\end{figure}

\begin{figure}[!htbp]
    \centering
    \includegraphics[width=\linewidth]{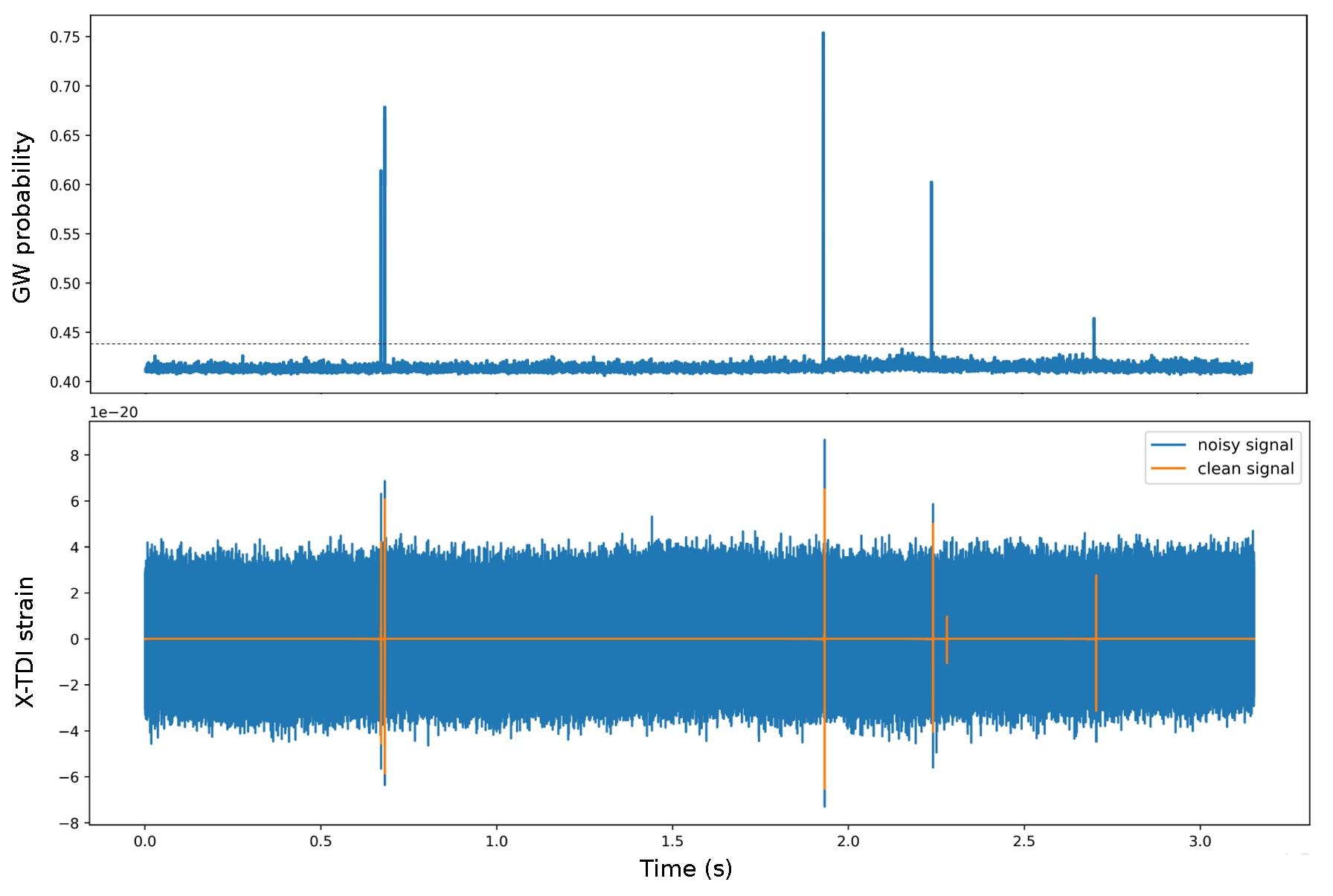}
    \caption{Comparison between the unblinded Sangria dataset (bottom) and our detection results (top). The horizontal line in the upper plot marks the GW probability threshold of 0.44, chosen to optimize merger detection. Detected mergers align with true events, though the smallest amplitude merger remains undetected.}
    \label{fig:result 2 vs unblind}
\end{figure}

Building on this pre-trained model, we proceeded to the primary objective of the Sangria LISA Data Challenge, applying the data processing procedures as described in Sec.\ref{sec:movingwindowsapproach}
A probability threshold of 0.44 was chosen as the minimum required to correctly identify mergers while minimizing false positives from noise.Applying this threshold (0.44) for GW merger detection, Fig.\ref{fig:result 2 vs unblind} shows five mergers identified in the Sangria blind dataset. The corresponding loss curve can be seen in Fig.\ref{fig:loss2}.

Comparison with the unblinded dataset (Fig.\ref{fig:result 2 vs unblind}) shows that detected mergers align with the true merger times, and detection probability correlates with merger amplitude. However, we can see that the smallest merger remains undetected, indicating that the QNN confuses GW mergers with such a small amplitude for noise.

\section{Discussion. QNN and classical NN results} \label{sec:discussion}

\begin{figure}
     \centering
     \includegraphics[width=\linewidth]{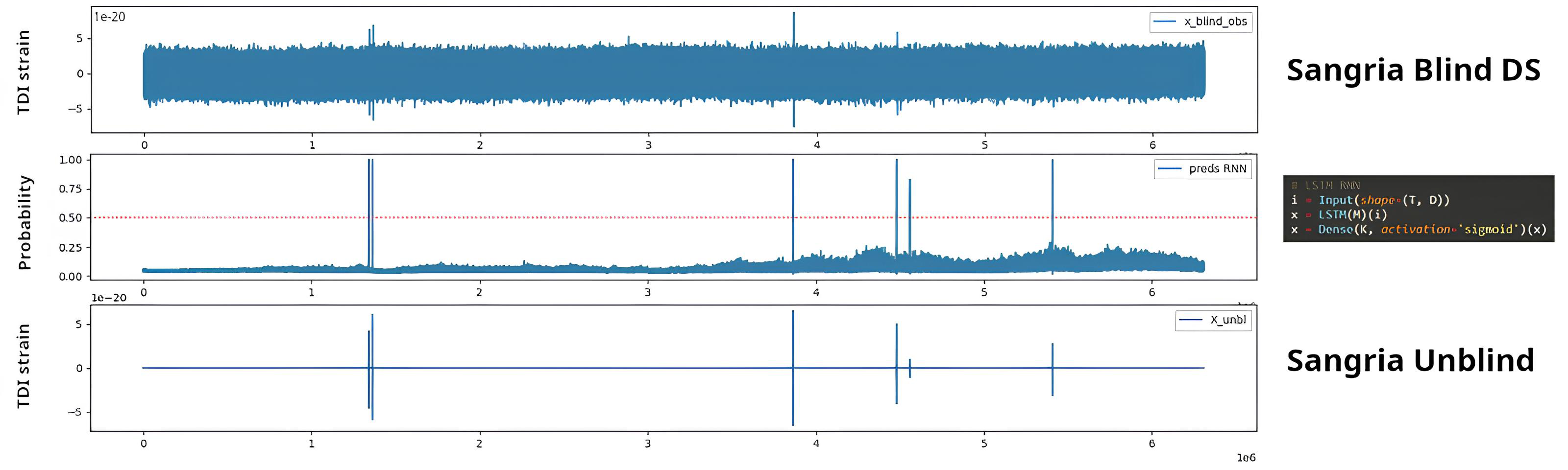}
     \caption{The Sangria blind dataset (up), the merger detection results using the RNN (middle) along with the model (right), and the unblinded Sangria dataset (down). By comparing the middle plot with the down plot, we can see that all 6 mergers were successfully identified. In the middle plot, it can be observed that the detection threshold is significantly below 0.5 (\url{https://indico.ict.inaf.it/event/2690/contributions/18574/attachments/8514/17535/Tonoiu-Daniel_flashtalk.pdf}) \cite{articolRNN}. Next to the middle plot, the RNN model implementation is shown.}
     \label{fig:rezultatRNN}
 \end{figure}

 Within our group, the Sangria LISA Data Challenge was solved using classical neural networks, specifically a recurrent neural network (RNN). This solution represents the first AI-based result within the LISA Consortium; all six mergers were successfully identified, with a GW detection threshold significantly below 0.5 (Fig.\ref{fig:rezultatRNN}). This threshold is somewhat lower than that obtained by the QNN. These results were presented as a poster at the \href{https://indico.ict.inaf.it/event/2690/}{International Conference on Machine Learning for Astrophysics 2nd Ed. - ML4ASTRO2} conference (\url{https://indico.ict.inaf.it/event/2690/contributions/18574/attachments/8514/17535/Tonoiu-Daniel_flashtalk.pdf}) and will be included in the paper \textit{GWEEP: A Deep Learning Toolkit for LISA Low-Latency Pipeline} (in preparation) \cite{articolRNN}. In parallel, several ML-based approaches for LISA data analysis have been proposed in the literature, demonstrating the growing interest in applying AI to GW astronomy \cite{LISApipelines2} \cite{LISApipelines4} \cite{LISApipelines5}.

 Classical NN and QNN characteristics, features and results are detailed in Table \ref{tab:classical_nn} and Table \ref{tab:qnn}. The RNN data is provided by five signal projections and one feature, totaling 6307180 samples, while one signal projection and four features totaling 126124 samples constitute the data for the QNN. Notably, the RNN’s label assignment differed slightly from the method described in Sec.~\ref{sec:movingwindowsapproach}: after labeling individual data points, each sample was assigned the label of the data point immediately following its window’s end. Both NNs use the same dataset for training (the Sangria train dataset), but the QNN needs 99.98\% less data samples than the classical NN, implying minimal pre-processing requirements. The QNN contains only 64 trainable parameters, the classical NN contains 4240385 trainable parameters, i.e. $\approx$88000\% more. The QNN model demonstrates exceptional data comprehension and generalization ability, while featuring a minimalist architecture.

 Regarding computational complexity, the classical NN scales as $O(2000N_{batch})$, increasing linearly with batch size $N_{batch}$. The QNN complexity scales as $O(n_{qubits}N_{samples})$ (note: the QNN batch size is always 1). The number of gates increases linearly with the qubit number $n_{qubits}$ and the number of samples $N_{samples}$. Both time complexities are linear.

\newcolumntype{C}[1]{>{\centering\arraybackslash}m{#1}}

\begin{table}[htbp]
    \centering
    \begin{adjustbox}{max width=\textwidth}
    \small
        \begin{tabular}{|C{0.5\linewidth}|C{0.5\linewidth}|}
            \cline{2-2} 
            \multicolumn{1}{c|}{} & \textbf{Classical NN} \\
            \hline
            \textbf{Signal projections} & X, Y, Z, A, E \\
            \hline
            \textbf{Sample features} & spectral entropy \\
            \hline
            \textbf{Total no. of samples} & 6307180 \\
            \hline 
            \textbf{Sample length} & 1000 \\
            \hline
            \textbf{Step size} & 100 \\
            \hline
            \textbf{Percent of samples used for training} & 68\% \\
            \hline
            \textbf{Percent of samples used for prediction} & 32\% \\
            \hline
            \textbf{Model} & Recurrent Neural Network \\
            \hline
            \textbf{No. of trainable parameters} & 4240385 \\
            \hline
            \textbf{Time complexity} & $O(2000N_{batch})$ \\
            \hline
            \textbf{Loss function} & binary cross entropy \\
            \hline
            \textbf{Detection threshold} & \textless0.5 \\
            \hline
        \end{tabular}
    \end{adjustbox}
    \caption{Summary of the classical NN used in the Sangria LISA Data Challenge. Similar to the approach taken with the QNN, the model architecture for these numbers was determined heuristically based on performance metrics obtained during training on the same dataset.}
    \label{tab:classical_nn}
\end{table}

\begin{table}[htbp]
    \centering
    \begin{adjustbox}{max width=\textwidth}
    \small
        \begin{tabular}{|C{0.5\linewidth}|C{0.5\linewidth}|}
            \cline{2-2} 
            \multicolumn{1}{c|}{} & \textbf{QNN} \\
            \hline
            \textbf{Signal projections} & A \\
            \hline
            \textbf{Sample features} & spectral entropy, power spectral density mean, standard deviation, maximum \\
            \hline
            \textbf{Total no. of samples} & 63062 \\
            \hline 
            \textbf{Sample length} & 100 \\
            \hline
            \textbf{Step size} & 10 \\
            \hline
            \textbf{Percent of samples used for training} & 50\% \\
            \hline
            \textbf{Percent of samples used for prediction} & 50\% \\
            \hline
            \textbf{Model} & QNN based on VQC \\
            \hline
            \textbf{No. of trainable parameters} & 64 \\
            \hline
            \textbf{Time complexity} & $O(n_{qubits}N_{samples})$ \\
            \hline
            \textbf{Loss function} & binary cross entropy \\
            \hline
            \textbf{Detection threshold} & 0.44 \\
            \hline
        \end{tabular}
    \end{adjustbox}
    \caption{Summary of the QNN used in the Sangria LISA Data Challenge}
    \label{tab:qnn}
\end{table}

\section{Conclusions} \label{sec:conclusions}

We developed a QNN based on the VQC algorithm, capable of distinguishing between noise and GW signals embedded in noise, as well as identifying GW mergers. The QNN was trained in two stages. First, it distinguished noise (0) from wave (1) samples. Second, it classified samples as non-merger (0) or merger (1). For the first stage, samples were extracted as adjacent, non-overlapping windows. For the second stage, overlapping moving windows were used. We computed features including spectral entropy, power spectral density statistics (mean, standard deviation), signal maximum, mean, and peak frequency. Feature selection was performed during training to optimize performance.
This results mark the first step of our approach towards building robust data analysis tools based on QNNs for fast detection and characterization of GW signals. Compared to the RNN, the QNN requires 99.98\% fewer data samples and uses only 0.00113\% of the trainable parameters. The QNN currently confuses the smallest amplitude merger with noise, whereas the RNN successfully detects it as a GW signal. To improve detection of low amplitude mergers, we plan to prepare an additional mock training dataset and test alternative ansatz designs. Yet, there is great potential that future QNN models will achieve such detection precision, featuring minimal data pre-processing routines and simplistic architectures.

Currently, the QNN focuses solely on GW detection. Future work will develop algorithms to identify GW source parameters after detection.

\section{Future work} \label{sec:future}

The next step toward detecting and characterizing all 6 GW mergers in the Sangria blind dataset is to design a QNN capable of identifying each merger within the time series. This effort builds on our previously established QNN. Possible scenarios include testing different ansatzes, preparing a new mock training dataset for further training, and further classical training of QNN outputs, as explored in hybrid quantum-classical models where quantum-generated features are optimized or classified using classical machine learning techniques \cite{hybrid1} \cite{hybrid2} \cite{hybrid3}. Following successful GW detection, the next phase will involve developing a dedicated algorithm to estimate the source parameters of the identified signals. 
All our work is done using local simulators of quantum computers. In order to prepare for the integration of quantum computers into our computing facilities, we plan to benchmark available devices using our QNN data analysis tools for GW detection. There are several established quantum computing hardware technologies: superconductivity, ion traps, neutral atoms and photonics.

\section{Acknowledgements}
This work is supported by the Romanian Ministry of Research, Innovation and Digitalization under the Romanian National Core Program LAPLAS VII - contract no. 30N/2023 and by ESA PRODEX project RoLISASpace.

\appendix
\section{Qubits, gates and circuits} \label{app:basicqc}

Qubits are the elementary units of quantum information. Unlike classical bits, they can exist in a superposition of the states '0' and '1', with complex amplitudes. The information stored into a qubit is manipulated through quantum logic gates, which can also carry information  from the classical world encoded into gate parameters. A quantum circuit consists of a sequence of quantum gates applied to one or more qubits, followed by a measurement. The measurement at the end of the circuit is necessary for retrieving the processed information and interpreting it \cite{sakurai}.

The state of a qubit is written as:

\begin{equation}
    |\phi\rangle = \alpha \ |0\rangle + \beta \ |1\rangle
\end{equation}

with $|\alpha|^2 + |\beta|^2 = 1$.

Quantum logic gates are operations that change the state of one or multiple qubits. Unlike classical operations, quantum operations are reversible. Quantum gates are represented as unitary matrices:

\begin{equation}
    U U ^\dag = U ^\dag U = I
\end{equation}

The most common gates are the superposition generating Hadamard gate $H$, the Pauli gates $X,Y,Z$, entanglement generating gates $CNOT, \ ZNOT$ that imply one or more control qubits and one target qubit. Some other useful gates are the rotation gates $R_x(\theta), \ R_y(\theta), \ R_z(\theta)$; $\theta \in [0,\pi]$ represents the rotation angle.

Visually, qubits are represented as horizontal lines. 
Each qubit is initialized with the state $|0\rangle$.

\begin{figure}[htbp]
    \centering
    \includegraphics[width=0.3\linewidth]{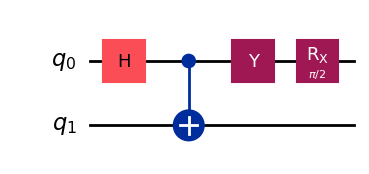}
    \caption{A series of gates applied to a circuit of two qubits: Hadamard, $CNOT$ with $q_0$ and control and $q_1$ as target, Pauli gate $Y$ and rotation around $x$ axis with angle $\pi /2$. This sequence of gates does not have any particular meaning.}
    \label{fig:gates}
\end{figure}

Gates are illustrated as symbols placed on the horizontal lines representing the qubits. Each gate symbol indicates a specific quantum operation applied to one or more qubits. Single-qubit gates are represented as boxes, while multi-qubit gates are depicted as dots on the control qubit lines and '$\bigoplus$' on the target qubit line that are connected with a vertical line (Fig.\ref{fig:gates}). The gates act from left to right; in our example (Fig.\ref{fig:gates}), the first operation performed is $H$, the second operation is $CNOT$ and so on. 

\begin{figure}[htbp]
    \centering
    \includegraphics[width=0.25\linewidth]{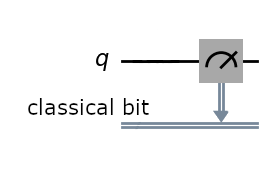}
    \caption{Measurement of a qubit symbol. The measurement result is stored in a classical bit.}
    \label{fig:meas}
\end{figure}

To access the result of the quantum circuit it is necessary to perform a measurement. This operation is typically illustrated as a meter-like icon (Fig\ref{fig:meas}). The measurement symbol is placed on the qubit lines corresponding to the qubits being measured, and it is usually connected with a vertical arrow to a classical bit double line to indicate that the result of the measurement is recorded in a classical register. 

\section{Quantum neural networks. Variational quantum classifiers} \label{app:basicqnn}

QNNs \cite{QNN} are the quantum counterpart of classical NNs. Several advantages are anticipated when using QNNs instead of classical NNs: requiring fewer training samples \cite{QNNsneedlesstrainingdata}, distinguishing between highly similar data classes with high accuracy \cite{QNNsdistinguishbetweensimilarclasses1}, \cite{QNNsdistinguishbetweensimilarclasses2}, better handling of noisy data \cite{QNNshandlenoisydatabetter1}, \cite{QNNshandlenoisydatabetter2}, and requiring fewer trainable parameters \cite{QNNSneedlesstrainingparams}.

Like classical NNs, QNNs can employ various architectures inspired by their classical counterparts. For classification tasks, a widely used QNN architecture is based on variational quantum circuits, similar to the Variational Quantum Classifier (VQC) algorithm \cite{parameterizedquantumcircuitsasqmlmodels, VQC}. 

The Variational Quantum Classifier (VQC) is a hybrid quantum-classical algorithm used for supervised learning tasks, such as classification. The main quantum circuit used in VQC is parameterized, meaning that it has trainable parameters (often rotation angles) that can be adjusted during training. This circuit, known as 'ansatz', serves as the model that maps input data to outputs. Classical inputs (e.g., feature vectors) are encoded into quantum states using a component called the feature map \cite{featuremapref1} \cite{featuremapref2}. This step is crucial because it determines how the quantum computer 'sees' the data. The feature map typically consists of quantum gates parameterized by classical data, encoding the information into the amplitudes and/or phases of the quantum state. Once the data is encoded, the parameterized ansatz circuit is applied to evolve the initial quantum state. \cite{vqa} \cite{parametizedquantumcircuits}. After processing by the ansatz, a measurement is performed to collapse the quantum state into classical outcomes, typically in the form of class probabilities. The measurement collapses the quantum state into a classical output, typically in the form of probabilities. These probabilities correspond to different classes in the classification problem and then they are processed classically to make a final prediction. A classical cost function (e.g., cross-entropy loss) evaluates the model's performance by comparing predicted and true labels. The parameters of the ansatz are optimized using classical algorithms such as Gradient Descent, COBYLA, or Adam. The optimization loop involves running the quantum circuit multiple times, each time adjusting the parameters slightly and computing the cost function. In our work, this type of QNN architecture is employed to address the Sangria LISA Data Challenge.

\bibliographystyle{plain}
\bibliography{bibliography}

\end{document}